\documentstyle[12pt,axodraw,epsfig]{article}

\catcode`\@=11
\long\def\@makefntext#1{ 
\protect\noindent \hbox to 3.2pt {\hskip-.9pt
$^{{\ninerm\@thefnmark}}$\hfil}#1\hfill} 

\def\thefootnote{\fnsymbol{footnote}}
\def\@makefnmark{\hbox to 0pt{$^{\@thefnmark}$\hss}}  

\def\ps@myheadings{\let\@mkboth\@gobbletwo
\def\@oddhead{\hbox{} 
\rightmark\hfil\ninerm\thepage}
\def\@oddfoot{}\def\@evenhead{\ninerm\thepage\hfil 
\leftmark\hbox{}}\def\@evenfoot{}
\def\sectionmark##1{}\def\subsectionmark##1{}}

\begin{document}


\newcommand{\symbolfootnote}{\renewcommand{\thefootnote}
        {\fnsymbol{footnote}}}
\renewcommand{\thefootnote}{\fnsymbol{footnote}}
\newcommand{\alphfootnote}
        {\setcounter{footnote}{0}
         \renewcommand{\thefootnote}{\sevenrm\alph{footnote}}}
\newcommand{\ba}{\begin{eqnarray}}
\newcommand{\ea}{\end{eqnarray}}
\newcommand{\crn}{\nonumber \\}





\newcounter{appendixc}
\newcounter{subappendixc}[appendixc]
\newcounter{subsubappendixc}[subappendixc]
\renewcommand{\thesubappendixc}{\Alph{appendixc}.\arabic{subappendixc}}
\renewcommand{\thesubsubappendixc}
        {\Alph{appendixc}.\arabic{subappendixc}.\arabic{subsubappendixc}}

\renewcommand{\appendix}[1] {\vspace{0.6cm}
        \refstepcounter{appendixc}
        \setcounter{figure}{0}
        \setcounter{table}{0}
        \setcounter{equation}{0}
        \renewcommand{\thefigure}{\Alph{appendixc}.\arabic{figure}}
        \renewcommand{\thetable}{\Alph{appendixc}.\arabic{table}}
        \renewcommand{\theappendixc}{\Alph{appendixc}}
        \renewcommand{\theequation}{\Alph{appendixc}.\arabic{equation}}
        \noindent{\bf Appendix \theappendixc #1}\par\vspace{0.4cm}}
\newcommand{\subappendix}[1] {\vspace{0.6cm}
        \refstepcounter{subappendixc}
        \noindent{\bf Appendix \thesubappendixc. #1}\par\vspace{0.4cm}}
\newcommand{\subsubappendix}[1] {\vspace{0.6cm}
        \refstepcounter{subsubappendixc}
        \noindent{\it Appendix \thesubsubappendixc. #1}
        \par\vspace{0.4cm}}



\def\abstracts#1{{
        \centering{\begin{minipage}{30pc}\tenrm\baselineskip=12pt\noindent
        \centerline{\tenrm ABSTRACT}\vspace{0.3cm}
        \parindent=0pt #1
        \end{minipage} }\par}}






\topsep=0in\parsep=0in\itemsep=0in
\parindent=1.5pc


\newcounter{itemlistc}
\newcounter{romanlistc}
\newcounter{alphlistc}
\newcounter{arabiclistc}
\newenvironment{itemlist}
        {\setcounter{itemlistc}{0}
         \begin{list}{$\bullet$}
        {\usecounter{itemlistc}
         \setlength{\parsep}{0pt}
         \setlength{\itemsep}{0pt}}}{\end{list}}

\newenvironment{romanlist}
        {\setcounter{romanlistc}{0}
         \begin{list}{$($\roman{romanlistc}$)$}
        {\usecounter{romanlistc}
         \setlength{\parsep}{0pt}
         \setlength{\itemsep}{0pt}}}{\end{list}}

\newenvironment{alphlist}
        {\setcounter{alphlistc}{0}
         \begin{list}{$($\alph{alphlistc}$)$}
        {\usecounter{alphlistc}
         \setlength{\parsep}{0pt}
         \setlength{\itemsep}{0pt}}}{\end{list}}

\newenvironment{arabiclist}
        {\setcounter{arabiclistc}{0}
         \begin{list}{\arabic{arabiclistc}}
        {\usecounter{arabiclistc}
         \setlength{\parsep}{0pt}
         \setlength{\itemsep}{0pt}}}{\end{list}}



\newcommand{\fcaption}[1]{
        \refstepcounter{figure}
        \setbox\@tempboxa = \hbox{\tenrm Fig.~\thefigure. #1}
        \ifdim \wd\@tempboxa > 6in
           {\begin{center}
        \parbox{6in}{\tenrm\baselineskip=12pt Fig.~\thefigure. #1 }
            \end{center}}
        \else
             {\begin{center}
             {\tenrm Fig.~\thefigure. #1}
              \end{center}}
        \fi}


\newcommand{\tcaption}[1]{
        \refstepcounter{table}
        \setbox\@tempboxa = \hbox{\tenrm Table~\thetable. #1}
        \ifdim \wd\@tempboxa > 6in
           {\begin{center}
        \parbox{6in}{\tenrm\baselineskip=12pt Table~\thetable. #1 }
            \end{center}}
        \else
             {\begin{center}
             {\tenrm Table~\thetable. #1}
              \end{center}}
        \fi}




\def\fnm#1{$^{\mbox{\scriptsize #1}}$}
\def\fnt#1#2{\footnotetext{\kern-.3em
        {$^{\mbox{\sevenrm #1}}$}{#2}}}


\font\twelvebf=cmbx10 scaled\magstep 1
\font\twelverm=cmr10 scaled\magstep 1
\font\twelveit=cmti10 scaled\magstep 1
\font\elevenbfit=cmbxti10 scaled\magstephalf
\font\elevenbf=cmbx10 scaled\magstephalf
\font\elevenrm=cmr10 scaled\magstephalf
\font\elevenit=cmti10 scaled\magstephalf
\font\bfit=cmbxti10
\font\tenbf=cmbx10
\font\tenrm=cmr10
\font\tenit=cmti10
\font\ninebf=cmbx9
\font\ninerm=cmr9
\font\nineit=cmti9
\font\eightbf=cmbx8
\font\eightrm=cmr8
\font\eightit=cmti8

\newcommand{\be}{\begin{equation}}
\newcommand{\ee}{\end{equation}}
\newcommand{\bea}{\begin{eqnarray}}
\newcommand{\eea}{\end{eqnarray}}
\newcommand{\al}{\alpha}
\newcommand{\Be}{\mbox{B}}
\newcommand{\gm}{\gamma}
\newcommand{\Gm}{\Gamma}
\newcommand{\dl}{\delta}
\newcommand{\Dl}{\Delta}
\newcommand{\eps}{\varepsilon}
\newcommand{\ep}{\varepsilon}
\newcommand{\kp}{\kappa}
\newcommand{\lm}{\lambda}
\newcommand{\Lm}{\Lambda}
\newcommand{\om}{\omega}
\newcommand{\pr}{\partial}
\newcommand{\pa}{\partial}
\newcommand{\dd}{\mbox{d}}
\newcommand{\dr}{{\rm d}}
\newcommand{\la}{\langle}
\newcommand{\ra}{\rangle}
\newcommand{\MS}{\mbox{MS}}
\newcommand{\up}{\underline{p}}
\newcommand{\uq}{\underline{q}}
\newcommand{\uk}{\underline{k}}
\newcommand{\um}{\underline{m}}
\newcommand{\uu}{\underline{u}}
\newcommand{\ual}{\underline{\alpha}}
\newcommand{\nn}{\nonumber}

\vspace{-1cm}
\noindent {\small BI-TP 97/13\\
           \small   MZ-TH/97-20\\[5mm]}
\begin{center}

\Large CALCULATION OF INFRARED-DIVERGENT FEYNMAN DIAGRAMS
WITH ZERO MASS THRESHOLD\footnote{Work supported by European project
Human Capital and Mobility under CHRX-CT94-0579}

\vspace{0.9cm}

\large J.~FLEISCHER$^a$\footnote{~E-mail:
fleischer@physik.uni-bielefeld.de},
~~V.A.~SMIRNOV$^b$\footnote{E-mail:
smirnov@theory.npi.msu.su}
\\
\baselineskip=13pt
\normalsize $^a$Fakult\"at f\"ur Physik, Universit\"at Bielefeld \\
\baselineskip=12pt
D-33615 Bielefeld , Germany \\
\baselineskip=13pt
$^b$Nuclear Physics Institute of Moscow State University \\
\baselineskip=12pt
119899 Moscow, Russian Federation \\[1cm]
\large
A.~FRINK$^c$\footnote{E-mail: frink@dipmza.physik.uni-mainz.de},
~~J.~K\"ORNER$^c$,
~~D.~KREIMER$^c$\footnote{Heisenberg Fellow},
~~J.B.~TAUSK$^c$,
~~K.~SCHILCHER$^d$\footnote{supported in part by Volkswagenstiftung,
permanent address: Inst.f.Physik, Univ., D-55099 Mainz}\\
\normalsize
$^c$Institut f\"ur Physik, Universit\"at Mainz \\
\baselineskip=12pt
D-55099 Mainz, Germany\\
$^d$ Dept.~of Physics, Univ.~of California,\\
San Diego, La Jolla, CA 92093-0319 USA\\[1cm]

\end{center}

\abstract{
Two-loop vertex Feynman diagrams with infrared and
collinear divergences are investigated
by two independent methods.
On the one hand, a method of calculating Feynman diagrams from their
small momentum expansion \cite{ft}  extended to diagrams with zero
mass thresholds \cite{fst} is applied. On the other hand, a
numerical method based on a two-fold integral representation is used
\cite{CKK}, \cite{FKKR}.
The application of the latter method is possible by using
lightcone coordinates in the parallel space. The numerical
data obtained with the two methods are in impressive agreement.}

\newpage

\section{Introduction}

The purpose of this paper is to calculate typical two-loop
IR-divergent vertex diagrams by two independent methods. One of
them is based on general formulae for asymptotic expansions of
Feynman integrals in momenta and masses \cite{ae,ae-proof} (see
\cite{vs} for a brief review) and subsequent use of conformal
mapping and summation by Pad\'{e} approximations \cite{ft,fst}.
The general simple formulae have been proven \cite{ae-proof}
at least for the Feynman integrals off the mass
shell\footnote{Note that for some typical on-shell limits
explicit formulae for asymptotic expansions have been
recently presented and applied in Ref. \cite{aems}.}.
However it is quite natural to expect that the same off-shell
formulae hold as well for the pure large mass limit when there
are no large momenta, as it has been confirmed in ref.~\cite{fst}.
This conjecture is likely to hold even for
Feynman integrals which possess IR and collinear divergences from
the very beginning.  To check the validity of this conjecture we
shall calculate IR-divergent diagrams~7 and~8 shown in Fig.~1
(the labelling is according to \cite{fst}, Fig.~3)
by use of the large mass expansion and compare results in a wide
range of external momenta with results based on
numerical integration.

The project is motivated by the demand to provide the necessary
integrals for the process $Z\rightarrow b\bar{b}$. The before-mentioned
divergences appear in the limit $m_b\to 0$, which is a natural limit
to take since $m_b$ is small compared to the other scales in this process.
The case of a finite $m_b$ mass can in principle also be handled
by both methods, see e.g. Ref. \cite{Rhei} for the momentum expansion
method applied to a finite diagram ($Case~5$, $m_5=m_6=M$, all others zero,
see Fig.~1). In this case, however, one obtains less
precise results because of the low thresholds, which are eliminated
in the present approach due to the factorization of logarithms.
For the numerical method the case $m_b\neq 0$ presents no difficulty.

The numerical methods, as introduced in \cite{CKK,FKKR,FKK,Laus} are
in the present paper extended to the case of
IR and collinear divergences. We will explain this extension below.
The two methods used in this paper are of complementary nature.
While for the method based on the large mass expansion the equal mass case
is simpler than the case of different masses, the situation
is reversed if we apply the two-fold integral
representation. The presence of poles of the type $\lim_{m_1\to m_2}
1/(m_1-m_2)$ in the integral representations
results in numerical instabilities which have to be handled
with due care.
Corresponding poles in the first method can be avoided from the very
beginning by calculating the Taylor coefficients analytically in terms
of rational numbers and $\zeta(2)$ for the equal mass case.
In the case of nonequal
masses the Taylor coefficients contain such poles of high order (see below),
which cancel in the $\lim_{m_1\to m_2}$. Numerically a close approach
is possible due to the use of a multiple precision package \cite{mpfun}.

The idea of the numerical approach is now well documented in the literature.
We will thus restrict ourselves to describe the necessary changes
demanded by the presence of IR and collinear divergences.
These changes are simple: one merely has to find subtraction terms
to absorb the divergences, which are sufficiently
easy to allow for a direct evaluation.
So there was no new conceptual
challenge present in the numerical method, but solely a technical
hurdle to be overcome.

The paper is organized as follows. First, we discuss the method
based on the large mass expansion
for $Cases~7$ and~$8$, considering the equal as well as different
masses in $Case~7$. We then discuss the relevant subtraction terms
for the numerical method. Once they are understood, all
the previous cases can be easily obtained.
We summarize the results in a set of tables, and compare
accuracies and CPU times for both methods.
We finally conclude that the agreement of data obtained by the two
independent methods confirms the usefulness of both.

\begin {figure} [htbp]
\begin{picture}(400,190)(-35,12)

\ArrowLine(35,100)(85,100)
\ArrowLine(35,150)(85,150)
\ArrowLine(35,100)(35,150)
\ArrowLine(85,150)(85,100)
\ArrowLine(85,100)(60,70)
\ArrowLine(60,70)(35,100)

\ArrowLine(60,50)(60,70)
\ArrowLine(35,150)(35,170)
\ArrowLine(85,170)(85,150)

\Text(15,135)[]{$k_2+p_1$}
\Text(107,135)[]{$k_2+p_2$}
\Text(60,160)[]{$k_2$}
\Text(60,45)[]{$p_1-p_2$}
\Text(60,25)[]{$generic$}
\Text(30,85)[]{$k_1+p_1$}
\Text(95,85)[]{$k_1+p_2$}
\Text(60,107)[]{$k_1-k_2$}
\Text(35,175)[]{$p_1$}
\Text(85,175)[]{$p_2$}

\Text( 25,118)[]{$m_3$}
\Text( 50, 70)[]{$m_1$}
\Text( 98,118)[]{$m_4$}
\Text( 75, 70)[]{$m_2$}
\Text( 60,140)[]{$m_5$}
\Text( 60, 87)[]{$m_6$}
\Vertex(35,100){2}
\Vertex(85,100){2}
\Vertex(35,150){2}
\Vertex(85,150){2}
\Vertex(60,70){2}


\DashLine(174,150)(224,150){3}
\Line(174,100)(224,100)
\DashLine(174,100)(174,150){3}
\DashLine(224,150)(224,100){3}
\Line(199,70)(174,100)
\Line(224,100)(199,70)


\Line(199,50)(199,70)
\DashLine(174,150)(174,170){3}
\DashLine(224,170)(224,150){3}

\Vertex(174,100){2}
\Vertex(224,150){2}
\Vertex(199,70){2}
\Vertex(174,150){2}
\Vertex(224,100){2}

\Text(199,25)[]{$Case\; 7$}


\DashLine(316,100)(316,150){3}
\DashLine(366,150)(366,100){3}
\DashLine(341,70)(316,100){3}
\DashLine(366,100)(341,70){3}


\DashLine(316,150)(316,170){3}
\DashLine(366,170)(366,150){3}
\Line(341,50)(341,70)
\DashLine(316,150)(366,150){3}
\Line(316,100)(366,100)

\Vertex(316,100){2}
\Vertex(316,150){2}
\Vertex(366,100){2}
\Vertex(366,150){2}
\Vertex(341,70){2}

\Text(341,25)[]{$Case\; 8$}


\vspace{-10pt}

\end{picture}
\caption {On-shell infrared divergent planar diagrams with zero thresholds
          (solid lines massive, dashed lines massless).}
\end{figure}
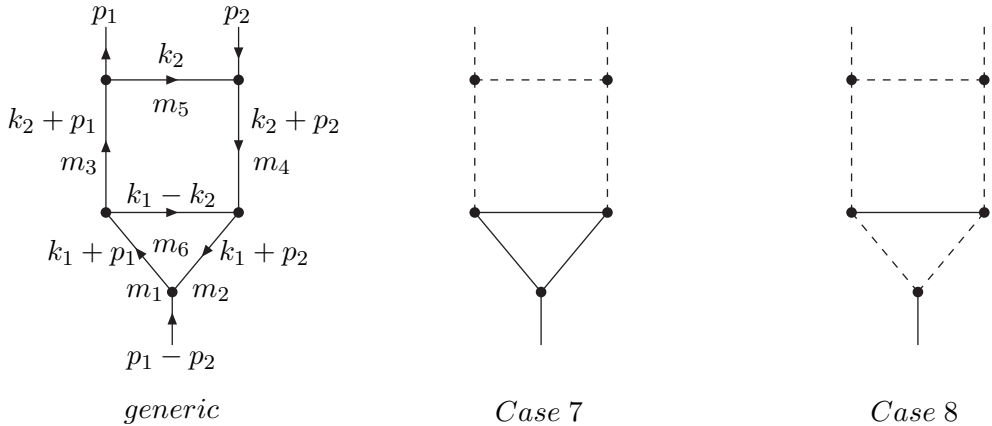


\section{\bf $Case~7$}
\subsection{Equal masses}

Let $F_{\Gamma} (p_1,p_2, M; \varepsilon)$ be the Feynman integral
corresponding to $Case~7$, with $m_3=m_4=m_5=0, \;
m_1=m_2=m_6=M, \; p_1^2=p_2^2=0, q=p_1-p_2$. We imply that the factor
$1/(k^2-m^2+i0)$ corresponds to a line, $1/(2\pi)^d$
is not included. For convenience we divide
our Feynman integrals by $i \pi^{d/2} \Gm(1+\eps) \mu^{-2\eps}$ per loop,
where $d=4-2\eps$ is the space-time dimension and $\mu$
the scale parameter of dimensional regularization \cite{dimreg}
(although we shall not usually write down $\mu$ explicitly).

Although the diagram has IR poles, up to $1/\eps^2$,
let us apply the general formula for the large mass
expansion \cite{ae,ae-proof} (as it was applied for similar
IR-finite planar diagrams in Ref.~\cite{fst}).
There are contributions from two subgraphs in this
formula: the graph $\Gm\equiv \gm_1$ itself and $\gm_2 = \{126\}$
(we denote subgraphs by collections of their lines).
The first, ``naive'', contribution is the formal Taylor expansion of the
initial diagram in external momenta. It reduces to two-loop vacuum graphs
with numerators, two massive and one massless line.
This term possesses IR divergences that were not present from
the very beginning.

The second contribution is nothing but the Taylor expansion of the heavy
triangle (with $p_1+k_2,p_2+k_2$ as external momenta, $k_2$ being the loop
momentum of the light triangle)
inserted into the light triangle. Now, this heavy triangle is a
scalar function of three variables: $(p_1 +k_2)^2, (p_2 +k_2)^2$ and
$q^2$. When performing its Taylor expansion, the factor
$(p_1 +k_2)^2$ leads to a cancellation of one of the lines in
the light triangle. This produces a one-loop massless diagram
with its external momentum on the lightcone. It is zero
within dimensional regularization. Thus we come to the conclusion
that only terms without such factors survive and
one can put $k_2$ to zero  in the expansion of the heavy triangle.
So, the term under consideration happens to be just a product
of two factors:
the light triangle with zero masses and the Taylor expansion
of the heavy triangle with $p_1,p_2, p_1-p_2$ as external momenta.
The light triangle is equal to
\be
- \frac{1}{\eps^2}
\frac{\Gm(1-\eps)^2}{\Gm(1-2\eps)}
\frac{1}{(-q^2-i0)^{1+\eps}} .
\label{lt}
\ee
(We shall later on omit $-i0$ in $-q^2-i0$, for brevity.)

The Taylor expansion of the heavy triangle is also calculated
explicitly:
\be
- \frac{1}{(M^2)^{1+\eps}} \frac{1}{\Gm(1+\eps)}
\sum_{n=0}^{\infty}
\frac{\Gm (n+1+\eps)}{2 (n+1) (2n+1)!!}
(q^2/ 2 M^2)^n .
\label{ht}
\ee

The second term does not involve UV divergences, in contrast
to the large mass expansions considered in Ref.~\cite{fst}.
In large mass expansions (as well as large momentum expansions off the mass
shell) induced IR divergences are usually cancelled by induced
UV divergences. Here we have an example of a new cancellation:
the induced IR divergences are cancelled by collinear
divergences that are present in the second term. To a large
extent this happens when we put to zero the above mentioned
one-loop massless diagram on the lightcone where the UV divergences are
cancelled by the collinear divergences.

Thus the contribution of the second subgraph takes the form
(with the normalization described above)
\bea
- 
\frac{1}{q^2} \frac{1}{M^2}
(-q^2/\mu^2)^{-\eps}
(M^2/\mu^2)^{-\eps}
\frac{1}{\eps^2}
\frac{\Gm(1-\eps)^2}{\Gm(1+\eps)\Gm(1-2\eps)}
\nn \\ \times
\sum_{n=0}^{\infty}
\frac{\Gm (n+1+\eps)}{2 (n+1) (2n+1)!!}
(q^2/ 2 M^2)^n .
\label{1s}
\eea

If an initial quantity is finite the cancellation of poles serves
as a good check of the asymptotic expansion. In our case, it is reasonable
to check that the poles in the sum of the terms in our expansion
are the same as
in the initial diagram.
To see that the poles in $\eps$ in the sum of the
two contributions are the same as in the initial Feynman integral
let us apply the following expression for the pole
part in $\eps$ of the product of three propagators, namely,
$1/k^2, 1/(k^2-2p_1 k)$ and $1/(k^2-2p_2 k)$ considered as a distribution
in $k$:
\be
i \frac{\pi^{2-\eps}}{(-q^2)^{1+\eps}}
\left\{ \left( -\frac{1}{\eps^2}
+ \gm_{\rm E} \frac{1}{\eps}
\right) \dl^{(d)} (k)
+  \frac{1}{\eps} \int_0^1 \frac{\dd z}{z} [ \dl(k-z p_1) + \dl(k-z p_2) -2\dl(k)]
 \right\} .
\label{pp}
\ee

It is easy to observe that
the term with $\dl^{(d)} (k)$ exactly corresponds to the pole
part of the second contribution.
Thus, the term with the integral over $z$ should
give the pole part of the naive contribution.
Let us therefore calculate the integral over $z$ of the rest of the diagram
(which is nothing but a one-loop triangle diagram with the external momenta
$(1-z) p_1, p_2 - z p_1$, plus a similar contribution, with $p_2$ instead
of $p_1$). Representing the result in the form of a Taylor expansion
in the external momenta we conclude that the pole part in $\eps$ of
the naive part is equal to
\be
\frac{1}{M^4} \frac{1}{\eps}
\sum_{n=0}^{\infty}  \frac{s_{n+1} (n+1)!}{2 (n+2) (2n+3)!!}
(q^2/ 2 M^2)^n ,
\label{npp}
\ee
where $s_n=S_1 (n)$ and we introduce
\be
S_k (n) = \sum_{j=1}^n \frac{1}{j^k}.
\label{skn}
\ee
This property has served as a check in our calculation.

The double pole in $\eps$ comes only from $\gm_2$. Using (\ref{1s})
we obtain the following explicit expression for the coefficient at
$1/\eps^2$:
\be
- \frac{2}{(q^2)^2} \arcsin^2 \sqrt{q^2/4M^2} \, .
\label{7-double_pole}
\ee

The naive part is calculated as described in \cite{fst}, see also \cite{ZiF}.
The two-loop bubble integrals with equal masses and one zero mass are again
of the type as used in \cite{fst}, i.e. expressible in terms of
$\Gamma$-functions as a result of which the Taylor coefficients of the naive
part are essentially rational numbers (apart from a $\zeta(2)$
as for the second contribution. These
Taylor coefficients have been calculated with FORM \cite{FORM} and the
Pad\'{e} approximants in turn by means of REDUCE \cite{REDUCE},
which allows an
easy change to floating point numbers of arbitrary precision. Adding up
the finite parts, the results are
presented in Table~1 with
the following normalization. In addition to
$-\pi^{d} \Gm(1+\eps)^2 \mu^{-4\eps}$, we extract the factor
$ \frac{\Gm(1-\eps)^2}{\Gm(1-2\eps)}\frac{1}{\mu^4}$, and we choose $\mu=M$.

\subsection{Two different masses}

Proceeding as before in $Case~7$ with two different masses,
$m_3=m_4=m_5=0, \; m_1=m_2=M_1, \; m_6=M_2$,
we come to the following result
for the contribution of the second subgraph:
\bea
\frac{1}{(-q^2)^{1+\eps}} \frac{1}{(M_2^2)^{\eps}}
\frac{1}{\eps^2}
\frac{\Gm(1-\eps)^2}{\Gm(1+\eps)\Gm(1-2\eps)}
\sum_{n=0}^{\infty} n! \left\{
\Gm (-n-1+\eps) \frac{(M_2^2)^{n+1}}{(M_2^2-M_1^2)^{2n+2}}
\right. \nn \\ \left.
+ \left( \frac{M_2^2}{M_1^2} \right)^{\eps} \;
\sum_{i=0}^{2n+1} \frac{(-1)^i \Gm (n-i+\eps)}{(2n-i+1)!}
\frac{(M_1^2)^{i-n}}{(M_2^2-M_1^2)^{1+i}} \right\} (q^2)^n .
\label{1sne}
\eea
Calculation of the pole and finite parts allows for the representation
\be
\frac{r_2}{\eps^2} + \frac{r_1 -r_2 L}{\eps}
+ r_0 -r_1 L + \frac{1}{2} r_2 L^2 \, ,
\label{pfp}
\ee
where $L = \ln M_2^2 + \ln(-q^2)$ and

\bea
r_0 = \frac{1}{2 M_1^2 q^2} \sum_{n=0}^{\infty} n!
(q^2/  M_1^2)^n
\nn \\
\times \left\{
\sum_{i=0}^{n-1} \frac{(n-i-1)!}{z^{i+1} (2 n - i + 1)!}
\left(\ln^2(1-z) - 2\zeta(2)
+2 \ln(1-z) s_{n - i-1} + s^2_{n - i-1} + t_{n - i-1} \right) \right.
\nn \\
 + \sum_{i=n}^{2n+1} \frac{(-1)^{i - n}}{3 z^{i+1} (i - n)! (2 n - i + 1)!}
\left( 3 \ln^2(1-z) s_{i - n}
\right. 
+3 \ln(1-z) (s^2_{i - n} - t_{i - n})
\nn \\
+ 3 s_{n + 1} t_{n + 1} + \ln^3(1-z) + s^3_{i - n} - s^3_{n + 1}
\left. \left.
+ u_{i - n} - u_{n + 1} -  3 s_{i - n} t_{i - n} \right) \right\} ,
\nn \\
r_1 = \frac{1}{M_1^2 q^2} \sum_{n=0}^{\infty} n!
(q^2/  M_1^2)^n
\left\{
\sum_{i=0}^{n-1} \frac{(n-i-1)!}{z^{i+1} (2 n - i + 1)!} (\ln(1-z)
+ s_{n -i -1})   \right. \nn \\
- \sum_{i=n}^{2n+1} \frac{(-1)^{i - n}}{2 z^{i+1} (i - n)! (2 n - i + 1)!}
\left(-\ln^2(1-z) -2 \ln(1-z) s_{i - n}
\right. \nn \\ \left. \left.
- s^2_{i - n} + t_{i - n} + s^2_{n + 1} - t_{n + 1} \right) \right\},
\nn \\
r_2 = \frac{1}{M_1^2 q^2} \sum_{n=0}^{\infty} n!
(q^2/  M_1^2)^n
\nn \\
\times \left\{
\sum_{i=0}^{n-1} \frac{ (n-i-1)!}{z^{i+1} (2 n - i + 1)!}
+ \sum_{i=n}^{2n+1} \frac{(-1)^{i - n}
\left(\ln(1-z) + s_{i - n} - s_{n + 1} \right)}{z^{i+1} (i - n)!
(2 n - i + 1)!}  \right\} , \nn
\eea

where $t_n = \zeta(2)-S_2(n)$ and $u_n = -2 \zeta(3) + S_3(n)$,
$S_k(n)$ given by Eq.~(\ref{skn}) and $z=1-M_2^2/M_1^2$. The
importance of presenting the above formulae lies in
their easy numerical evaluation. If instead we would use the
expanded form, 30 coefficients could not even be compiled
anymore simultaneously. 

We also calculate (in the same way as in the case with equal masses)
a quantity, the pole part of which is equal to the pole part of the
naive contribution:

\newpage

\bea
\frac{1}{q^2} \frac{1}{\eps}
\sum_{n=1}^{\infty} n! \sum_{j=1}^{n} \frac{1}{j}
\nn \\ \times
\left\{
(-1)^j  \sum_{i=0}^{j}
{2n-j+i+1 \choose 2n-j+1}
\frac{\Gm (j-i-n-1+\eps)}{(j-i)!}
\frac{(M_2^2)^{n+i-j+1-\eps}}{(M_2^2-M_1^2)^{2n-j+i+2}}
\right. \nn \\ \left.
+ \sum_{i=0}^{2n-j+1}  {j+i \choose j}
\frac{(-1)^i \Gm (n-j-i+\eps)}{(2n-j-i+1)!}
\frac{(M_1^2)^{i+j-n-\eps}}{(M_2^2-M_1^2)^{j+i+1}}
\right\} (q^2)^n .
\label{0snepp}
\eea
Explicitly, this gives
\bea
 \frac{1}{\eps} \frac{2}{M_1^2 q^2} \sum_{n=1}^{\infty} n! (q^2/M_1^2)^n
\nn \\
\times \sum_{j=1}^{n} \frac{(-1)^j}{j}
\left\{
 (-1)^n  \sum_{i=0}^{j} {2n-j+i+1 \choose 2n-j+1}
\frac{s_{n-j+i+1} (1-z)^{n-j+i+1}}{(j-i)!(n-j+i+1)!z^{2n-j+i+2}}
\right. \nn \\ \left.
+  \sum_{i=0}^{n-j-1}  {j+i \choose j} \frac{(n-j-i-1)!}{(2n-j-i+1)!z^{j+i+1}}
+ \sum_{i=0}^{n+1} (-1)^i {n+i \choose j}
\frac{\ln(1-z) + s_{i}}{i!(n+1-i)!z^{n+i+1}} \right\} \, .
\eea

Furthermore this representation of the pole part has been used
as a check of the naive part.

Finally there remains to calculate the finite contribution of the naive
part. Again we proceed as in \cite{fst} and \cite{ZiF}. The difficulty
arising now is a complication in the bubble integrals with two different
masses and one zero mass (e.g., $m_3=0$):

\begin{equation}
V_B(\alpha,\beta,\gamma,m_1,m_2,m_3)=\frac{(-1)^{(\alpha+\beta+\gamma)}}
{(i {\pi}^{\frac{d}{2}})^2} \int
\frac{d^dk_1 d^dk_2}{(k_1^2-m_1^2)^{\alpha}(k_2^2-m_2^2)^{\beta}
                                           ((k_1-k_2)^2-m_3^2)^{\gamma}} ,
\label{VBs}
\end{equation}

Apart from $\Gamma$-functions these contain now the hypergeometric function
${~}_2F_1(a,b,c;z)$ with the argument $z=1-\frac{m_1^2}{m_2^2} \le 1$
\cite{Ber}
($z=0$ and ${~}_2F_1=1$ in the equal mass case!). Therefore it does not seem
to be advisable anymore to use FORM for the evaluation of the Taylor
coefficients since too complicated expressions would arise. Instead we
extended a program developed for arbitrary nonzero masses \cite{Rhei} based
on the multiple precision FORTRAN by D.H. Bailey \cite{mpfun}. In this
framework the occurrence of a zero mass, i.e. the additional IR divergences
cause the following complication: while for arbitrary nonzero masses only the
bubble integrals

\begin{equation}
V_B(1,1,n,m_1,m_2,m_3)=F(1,1,n,m_1,m_2,m_3)+\frac{1}{\varepsilon}~
\frac{1}{(n-1)(n-2)}\frac{1}{(m^2_3)^{\left( n-2 \right)}}, ~n > 2 ,
\label{poleps}
\end{equation}
(F($\cdots$) being the finite part) are divergent like $\frac{1}{\varepsilon}$
(apart from $V_B(1,1,1)$ and $V_B(1,1,2)$ which have also
$\frac{1}{{\varepsilon}^2}$
-terms), in the present case all the needed bubble integrals
$V_B(1,m,n)$ with $m \ge 1$
and $n \ge 3$ have a $\frac{1}{\varepsilon}$-term, which is to be calculated
by means of the hypergeometric function ${~}_2F_1(a,b,c;z)$ with integer
indices $a,b,c$ ($d=4$) and $c=a+b+1$. This ${~}_2F_1$ can be expressed
in terms of $ln(1-z)$ and powers of $z$ and $1-z$.

Knowing the divergent parts of the bubble integrals explicitly,
the recursion for their finite part can be performed and the complete
contribution to the finite part of the diagram can be calculated.

Finally two further remarks are in order. The first concerns the evaluation
of the second subgraph. As one observes from the Taylor series of $r_0, r_1$
and $r_2$ in (9), these contain high powers of $\frac{1}{z}$.
In the equal mass
case ($z=0$), the expansion of the $ln(1-z)$ to high powers cancels all
inverse powers of $z$ and the finite coefficients as described in
Sect.~2.1 are obtained. Numerically, however, due to the use of the multiple
precision program of Bailey \cite{mpfun} for mass ratios close to 1 (e.g.
$z \sim 10^{-3}$) still stable results in close agreement with the equal
mass case can be obtained.

The second remark concerns adding up of the various
contributions.  There are two possibilities: one can sum the
different series ( for $r_0, r_1$ etc. ) by means of Pad\'{e} 's,
multiply the various results with the corresponding powers of $L$
(see e.g. (9)) and sum --- or one can add up on the level of the
coefficients, multiplying the Taylor coefficients with the powers
of $L$, and then apply Pad\'{e} 's. We used the latter procedure
since the coefficients are known to high precision (of the order
of 50 decimals). The summed series were of course by far not that
precise and if cancellations occur, one looses more precision
than necessary. That in this manner correct results are obtained
is demonstrated by comparing the results with those by the numerical
method (see tables).

While in the present paper adding different Taylor series with kinematical
factors is performed for the calculation of one diagram only, the same
procedure will also work when adding scalar amplitudes from various Feynman
diagrams - and may indeed work in an optimal way.

For the pole parts we do not show tables since they are either
given explicitly in terms of known functions or arbitrarily many
coefficients can be obtained easily and thus arbitrarily high
precision.  The results for the finite part are presented in
Table~2.  We use the same normalization as in the case with the
equal masses and choose $\mu=M_1$.

\section{\bf $Case~8$}

Let here $F_{\Gamma} (p_1,p_2, M; \varepsilon)$ be the Feynman integral
corresponding to $Case~8$, with $m_i =0,\; i=1, \ldots , 5, \;m_6=M$.
Now, in its large mass expansion, we have contributions from
four subgraphs: $\gm_1\equiv\Gm,\gm_2=\{3456\}, \gm_3=\{126\}$,
and $\gm_4=\{6\} $.
In the last two contributions, the factor (\ref{lt}) is again naturally
factorized. Straightforward  calculation leads to the following result
for the sum of them:
\bea
- \frac{1}{(M^2)^{1+\eps}} \frac{1}{(-q^2)^{1+\eps}}
\frac{1}{\eps^3} \frac{ \Gm(1-\eps)^3}{\Gm(1-2\eps)}
\nn \\ \times \left\{
\sum_{n=0}^{\infty} \frac{n!}{\Gm (n+2-\eps)}
(-q^2/  M^2)^n - (-q^2/ M^2)^{-\eps}
\sum_{n=0}^{\infty} \frac{\Gm (n+1-\eps)}{\Gm (n+2-2\eps)}
(-q^2/  M^2)^n \right\} .
\label{g34}
\eea

The calculation of the contribution of $\gm_2$ is similar to
the corresponding calculations for Cases~1 and~5 in ref.~\cite{fst}, with
the following result:
\be
\frac{1}{(M^2)^{2+\eps}}
\frac{1}{(-q^2)^{\eps}}
\frac{1}{\eps}
\sum_{n=0}^{\infty} c^{(2)}_n (\eps) (q^2/M^2)^n ,
\label{C2}
\ee
where
\bea
c^{(2)}_n (\eps) =
\sum_{{i_1,i_2,n_3 \geq 0,\; i_1+i_2+n_3 \; \mbox{\scriptsize even}}
\atop {i_1+i_2+n_3 \leq 2n}} \sum_{j_3 \geq 0}^{(n_3,-2)}
(-1)^{(i_1+i_2+n_3)/2} \;\frac{(n-(i_1+i_2-n_3)/2)!}{(n-(i_1+i_2+n_3)/2)!}
\nn \\
\times \frac{i_1!i_2! \theta (i_1+i_2-j_3)
\theta (i_1-i_2+j_3) \theta (-i_1+i_2+j_3)}{((n_3-j_3)/2)!
((i_1+i_2-j_3)/2)! ((i_1-i_2+j_3)/2)! ((-i_1+i_2+j_3)/2)!}
\nn \\
\times
\frac{\Gm(1-\eps)
\Gm((i_1+i_2-j_3)/2+1-\eps)}{\Gm(1+\eps)\Gm((i_1+i_2-j_3)/2+2-2\eps)}
\nn \\
\times C(1+i_2, 3+n+(i_1-i_2+n_3)/2; (i_1+i_2+n_3)/2),
\label{Case8_2}
\eea
and
\be
C(r_1, r_2; s) =
\frac{\Gm(r_1+r_2 -s-\frac{d}{2})\Gm(s-r_2
+\frac{d}{2})}{\Gm(r_1)\Gm(s+\frac{d}{2})} .
\ee
Note that this expression is obtained from the corresponding contribution
of $Case~5$ in \cite{fst} by the change
\bea
C(2+i_1 +i_2, 2+n-(i_1+i_2-n_3)/2; (i_1+i_2+n_3)/2)
\nn \\ \to
C(1+i_2, 3+n+(i_1-i_2+n_3)/2; (i_1+i_2+n_3)/2) . \nn
\eea

To see that the poles in $\eps$ in the sum of all the
four contributions are the same as in the initial Feynman integral
let us once again apply eq.~(\ref{pp}). Now we observe that
the term with $\dl^{(d)} (k)$ corresponds to the pole
part of the sum of the contributions from $\gm_3$ and $\gm_4$
given by (\ref{g34}). Thus, the term with the integral over $z$ should
give the pole part of the sum of contributions from $\gm_1$ and $\gm_2$.
So, we perform integration over $z$ of the one-loop triangle
diagram with one non-zero and two zero masses. Furthermore we apply
general formulae for the large mass expansion to this very triangle and
eventually conclude that the pole part in $\eps$
of the sum of the contributions from $\gm_1$ and $\gm_2$ should be equal to
the pole part of the following quantity:
\bea
\frac{1}{q^2} \frac{1}{(M^2)^{1+\eps}}
\frac{2}{\eps \Gm(1+\eps)} \sum_{n=1}^{\infty}  n! \left( \sum_{j=1}^n
\frac{C(j+1, 2n-j+2;n)}{j} \right) (q^2/ M^2)^n
\nn \\
+ 2\frac{1}{(-q^2)^{1+\eps}} \frac{1}{M^2}
\frac{\Gm(1-\eps)}{\eps^2}
\sum_{n=1}^{\infty} \frac{s_n \Gm(n+1-\eps)}{\Gm(n+2-2\eps)}
(-q^2/ M^2)^n .
\label{12pp}
\eea
Note that double poles in (\ref{12pp}) cancel so that
there must be
as well a cancellation of double poles in the sum of the
contributions of $\gm_1$ and $\gm_2$. Again this has been checked
and also that the single pole part agrees with the one obtained
by direct calculation of the sum of $\gm_1$ and $\gm_2$.

For $Case~8$, the double pole in $\eps$ originates only from the sum of
the contributions of $\gm_3$ and $\gm_4$. From (\ref{g34}) we have
the following formula for its coefficient:
\be
\frac{1}{(q^2)^2} \left\{ \ln (-q^2/M^2) \ln (1+q^2/M^2)
+ \mbox{Li}_2 (-q^2/M^2) \right\} \, .
\label{8-double_pole}
\ee

Concerning the finite part, the situation is similar as in Sect. 2.1,
$Case~7$, with equal masses. The bubble integrals are of the same type
as in $Case~5$ \cite{fst} and the calculation of the Taylor coefficients
have been done with FORM and the Pad\'{e} 's with REDUCE. Again we
present no data for the divergent parts since arbitrary precision
can easily be obtained from the expansion.
The results for the finite part are presented in Table~3
where the same normalization as in $Case~7$ is used.

\section{The Numerical Method}
The two-fold integral representation derived in \cite{2lp3pt,CKK,FKK}
cannot be naively applied to cases infected by genuine IR or collinear
divergences. Nevertheless, all these divergences  can be
handled by appropriate subtraction terms. One observes
a few new characteristics in such cases:
\begin{itemize}
\item
the divergences are most easily absorbed using lightcone coordinates
for internal momenta in the parallel space,
\item
some of the remaining domains of integration
are unbounded, cf.~Fig.(5).
\end{itemize}
We will comment on these features in the next three subsections.

\newcommand{\eq}[1]{Eq.~(\ref{eq:#1})}
\newcommand{\fig}[1]{Fig.~\ref{fig:#1}}
\renewcommand{\O}{{\cal O}}

\subsection{Preparations}

The graph we have to calculate is
\begin{eqnarray}
I & = & \mu^{2d-12} \int d^d k d^d l \frac{1}{P_1 P_2 P_3 P_4 P_5 P_6}
\nonumber\\
& = & \mu^{2d-12} \int d^d k \frac{1}{P_4 P_5 P_6} C(k) \, ,
\end{eqnarray}
where $C(k)$ denotes the inner one-loop triangle graph with external
momenta shifted by $k$ (\fig{oneloop}).

\begin{figure}[ht]
\begin{center}
\epsfig{file=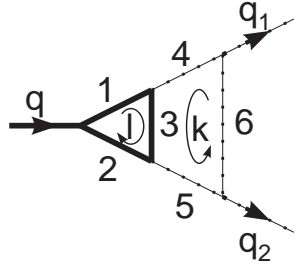,width=0.25\linewidth}
\end{center}
\caption{Propagator numbering, external and loop momenta}
\label{fig:graph}
\vspace*{1cm}
\end{figure}

\begin{figure}[ht]
\begin{center}
\epsfig{file=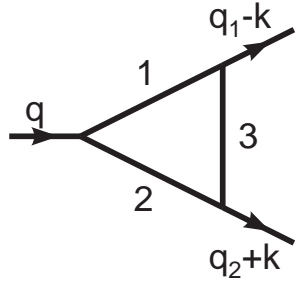,width=0.25\linewidth}
\end{center}
\caption{Oneloop subgraph $C(k)$}
\label{fig:oneloop}
\vspace*{1cm}
\end{figure}

For our calculation, we choose the rest frame of the decaying particle
and the outgoing particles moving along the $x$ axis. This is the
natural reference frame for parallel-/orthogonal space splitting.
Together with the condition that both outgoing momenta are on the
lightcone, the external momenta can be expressed through a single
parameter $e$ (cf. \fig{graph}):
\begin{eqnarray}
q = q_1+q_2 & = & (2e,0,\vec{0})
\nonumber\\
q_1 & = & (e,e,\vec{0})
\nonumber\\
q_2 & = & (e,-e,\vec{0})
\end{eqnarray}
As usual we parameterize the loop momenta $l$ and $k$ as follows:
\begin{equation}
l^\mu=(l_0,l_1,\vec{l}_\perp)
\quad\mbox{and}\quad
k^\mu=(k_0,k_1,\vec{k}_\perp) \, ,
\end{equation}
further we define
\begin{equation}
s=l_\perp^2 \quad\mbox{and}\quad t=k_\perp^2 \, .
\end{equation}
In contrast to \cite{CKK,FKK} we apply a more symmetric
substitution to linearize the propagators in $l_0$, $l_1$, $k_0$
and $k_1$ (the Jacobian gives a factor 2 for each loop momentum):
\begin{eqnarray}
l_0 & \to & l_0 + l_1
\nonumber\\
l_1 & \to & l_1 - l_0
\nonumber\\
k_0 & \to & k_0 + k_1
\nonumber\\
k_1 & \to & k_1 - k_0 \, ,
\end{eqnarray}
which is equivalent to expressing the propagators in lightcone
variables from the beginning. After this substitution the propagators
become
\begin{eqnarray}
P_1 & = & 4 l_0 (l_1+e) - s - m_1^2 + i0
\nonumber\\
P_2 & = & 4 l_1 (l_0-e) - s - m_2^2 + i0
\nonumber\\
P_3 & = & 4 (l_0+k_0)(l_1+k_1) - s - t - \sqrt{s}\sqrt{t}z - m_3^2 + i0
\nonumber\\
P_4 & = & 4 k_0 (k_1-e) - t - m_4^2 + i0
\nonumber\\
P_5 & = & 4 k_1 (k_0+e) - t - m_5^2 + i0
\nonumber\\
P_6 & = & 4 k_0 k_1 - t - m_6^2 + i0
\end{eqnarray}
Since we are interested in the collinear divergent case, we now set
$m_4$ and $m_5$ equal to 0, $m_6$ should be kept small for the {\em finite}
contribution as a regulator. The limit will be made later analytically.
The calculation is valid for arbitrary masses $m_1$, $m_2$ and $m_3$,
except for the trivial case $m_1=m_2=m_3=0$.

In this representation the collinear divergence along the lines
$k_0=0$ and $k_1=0$ for $t=0$ is obvious. To cure it, we adopt the
following subtraction scheme for $C(k)=C(k_0, k_1, t)$:
\begin{equation}
\begin{array}{rclr}
C(k_0,k_1,t) & = & \hphantom{+}
             C(k_0,  k_1, t) - C(k_0, 0, 0)
           - C(0, k_1, 0) + C(0, 0, 0)
& (I)
\\
& & + C(k_0, 0, 0) - C(0, 0, 0)
& (II)
\\
& & + C(0, k_1, 0) - C(0, 0, 0)
& (III)
\\
& & + C(0, 0, 0) \, .
& (IV)
\end{array}
\end{equation}
Contribution (I) will be finite and can be calculated in $d=4$
dimensions, since we subtracted out the collinear
divergences and added again the twice subtracted infrared
divergence at $k \equiv 0$. Contributions (II) and (III) contain
the collinear divergence which starts at $1/\eps$, and (IV) is the
overall infrared divergence with a $1/\eps^2$ pole. Therefore these
have to be calculated in $d=4-2\eps$ dimensions.

\subsection{Divergent parts}

Contribution (IV) is easy: it is simply a product of two one-loop
diagrams, where the $k$ loop is completely massless:
\begin{equation}
I_{\rm IV} = -i \pi^{d/2}
             \frac{\Gamma^2\left(\frac{d-4}{2}\right)
                   \Gamma\left(\frac{6-d}{2}\right)}
                  {\Gamma\left(d-3\right)}
             (-q^2-i0)^\frac{d-6}{2}
             C(0,0,0) \, .
\end{equation}
The expansion of the $\Gamma$ functions starts with $1/\eps^2$, so
$C(0,0,0)$ has to be calculated up to $\O(\eps^2)$. The last order
is done numerically.

Contribution (II) is more complicated, since the $l$ and $k$ integrations
do not decouple. However, the integrand is independent of $z$, the
angle between $\vec{l}_\perp$ and $\vec{k}_\perp$, so there is no
function with a cut in the complex $k_1$ plane. We have
\begin{eqnarray}
I_{\rm II} & = & \frac{\pi^\frac{d-2}{2}}
                      {2\Gamma\left(\frac{d-2}{2}\right)}
                 \int\limits_{-\infty}^{+\infty} dk_0
                 \big( C(k_0,0,0) - C(0,0,0) \big)
                 \int\limits_0^\infty t^\frac{d-4}{2} dt
                 \int\limits_{-\infty}^{+\infty} dk_1
                 \frac{1}{P_4 P_5 P_6} \, ,
\end{eqnarray}
and perform the $k_1$ integration with Cauchy's theorem.
We close the contour in the lower half plane, and
there is only a contribution from $P_5$ in the finite interval
$0 < k_0 < e$, outside this interval all poles are on the same side
of the real axis.

The integration over $t$ is of the form
\begin{equation}
\int\limits_0^\infty dt \frac{t^\frac{d-6}{2}}{t+4k_0(k_0+e)}
=
(4k_0(k_0+e))^\frac{d-6}{2} \Gamma\left(\frac{d-4}{2}\right)
                            \Gamma\left(\frac{6-d}{2}\right) \, ,
\end{equation}
so after a substitution $k_0=-z e$ the result is
\begin{eqnarray}
I_{\rm II} & = & -i \pi^{d/2}
                 \frac{\Gamma\left(\frac{d-4}{2}\right)
                       \Gamma\left(\frac{d-6}{2}\right)}
                      {\Gamma\left(\frac{d-2}{2}\right)}
                 (-q^2-i0)^\frac{d-6}{2}
\nonumber\\
& &              \int\limits_0^1 dz
                 \frac{1}{z} \left[z(1-z)\right]^\frac{d-4}{2}
                 \left( C(ze,0,0) - C(0,0,0) \right) \, .
\end{eqnarray}
The expansion of the $\Gamma$ functions starts at $1/\eps$, so
the integral over $z$, which will be done numerically, has to be
evaluated up to $\O(\eps)$.

Contribution (III) can be handled analogously by performing the
$k_0$ integration with Cauchy's theorem and gives
\begin{eqnarray}
I_{\rm III} & = & -i \pi^{d/2}
                  \frac{\Gamma\left(\frac{d-4}{2}\right)
                        \Gamma\left(\frac{d-6}{2}\right)}
                       {\Gamma\left(\frac{d-2}{2}\right)}
                  (-q^2-i0)^\frac{d-6}{2}
\nonumber\\
& &               \int\limits_0^1 dz
                  \frac{1}{z} \left[z(1-z)\right]^\frac{d-4}{2}
                  \left( C(0,ze,0) - C(0,0,0) \right) \, .
\end{eqnarray}
For the symmetric case $m_1=m_2$, $I_{\rm III}$ is equal to $I_{\rm II}$.

\subsection{The finite part}

The most difficult is the finite part from contribution (I).
In principle, the calculation is the same as in \cite{CKK,FKK},
but care has to be taken due to the divergent behavior of the individual
parts of this contribution.
The integration over $z$, the angle between $\vec{l}_\perp$ and
$\vec{k}_\perp$, is elementary for each for the four terms.
In fact, only the first term with the full $C(k_0,k_1,t)$ dependence
depends also on $z$, and therefore gives a result with a cut in
the complex $l_1$ and $k_1$ plane.

\begin{figure}[ht]
\begin{center}
\epsfig{file=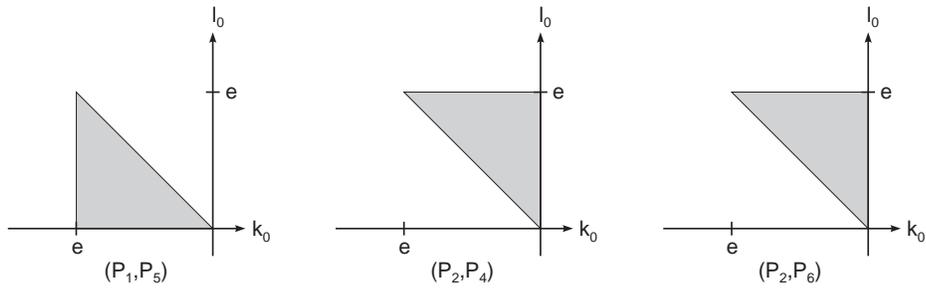,width=0.8\linewidth}
\end{center}
\caption{Contributions from terms 1 and 2}
\label{fig:col1}
\vspace*{1cm}
\end{figure}

The $l_1$ and $k_1$ integrations are done with the aid of the
residue theorem. Each term is still convergent on its own,
so the contours can be closed independently.
For the $C(k_0,k_1,t)$ term we have to take into account the cut
which is in the upper half plane if $l_0+k_0<0$ and in the lower
half plane if $l_0+k_0>0$. By checking the half plane where the
propagators have their poles, we find three contributing triangles
in the $(l_0,k_0)$ plane from the pairs $(P_1,P_5)$, $(P_2,P_4)$ and
$(P_2,P_6)$, depicted in \fig{col1}.

For the second term, $C(k_0,0,0)$, we can close the contours
arbitrarily. If we choose to close them in the same way as above
(this is equivalent to avoiding poles from $P_3$, the propagator that
depends on both loop momenta) we find the same
triangles from the same propagators as above.

\begin{figure}[ht]
\begin{center}
\epsfig{file=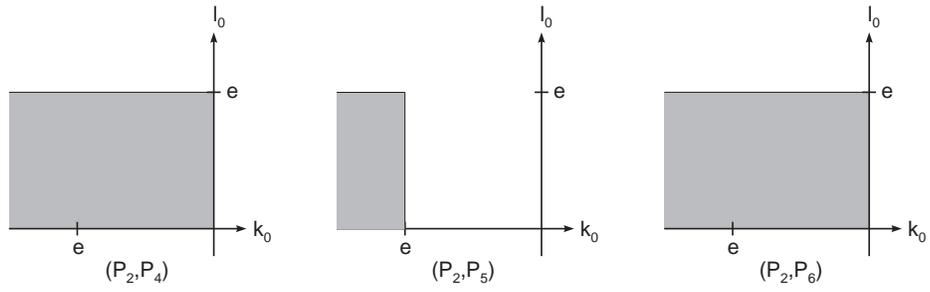,width=0.8\linewidth}
\end{center}
\caption{Contributions from term 3}
\label{fig:col2}
\vspace*{1cm}
\end{figure}

With $C(0,k_1,0)$ we find something new in our parallel/orthogonal
space technique: regardless how we close the contours, we end up
with an unbounded area in the $(l_0,k_0)$ plane. If we continue with
the strategy of avoiding poles from $P_3$, we have to close the $l_1$
contour in the upper half plane if $l_0>0$ and in the lower half
plane if $l_0<0$, furthermore we close the $k_1$ contour always in
the upper half plane. This gives us the three areas shown in
\fig{col2} from $(P_2,P_4)$, $(P_2,P_5)$ and $(P_2,P_6)$ respectively.

\begin{figure}[ht]
\begin{center}
\epsfig{file=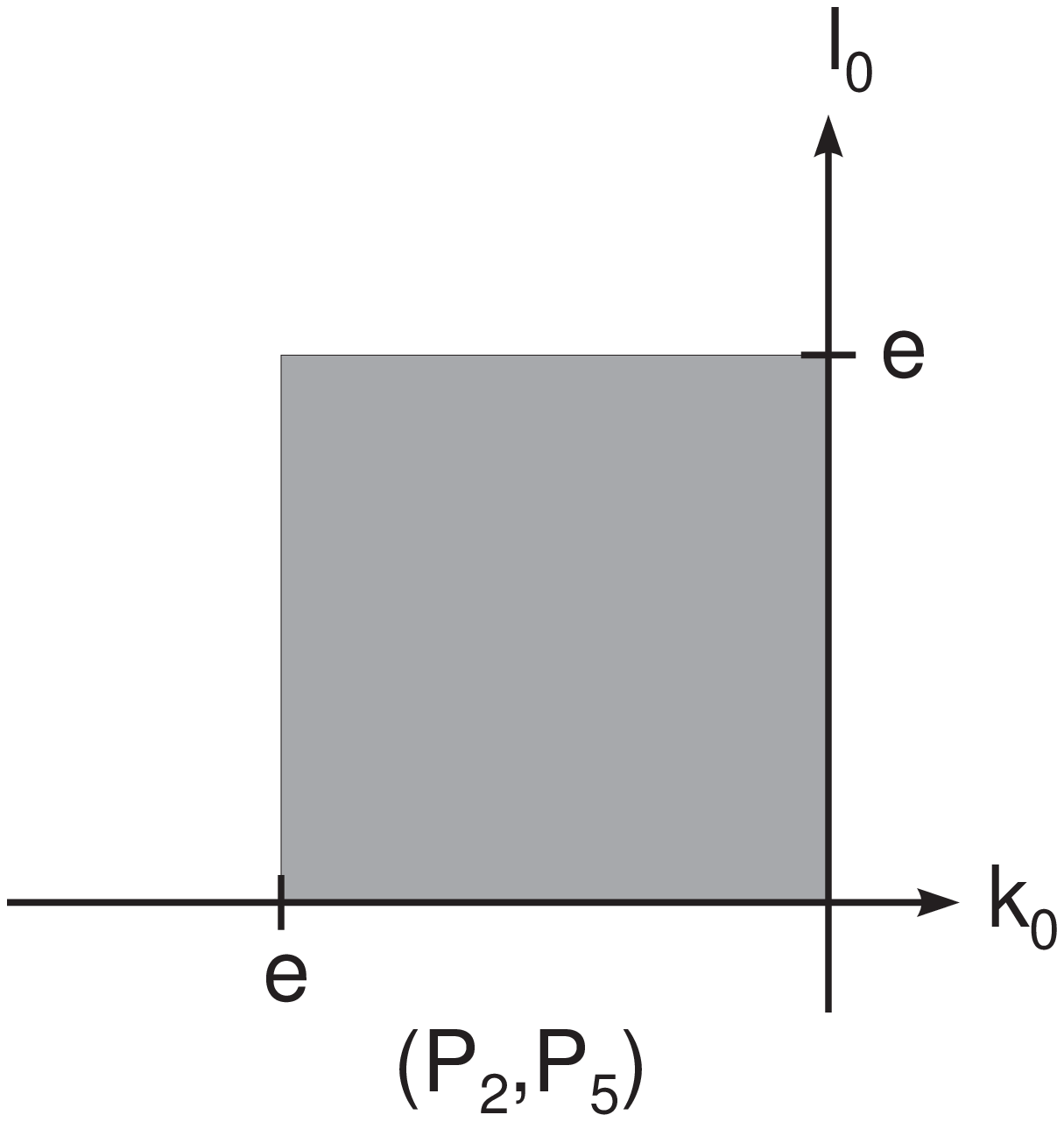,width=0.25\linewidth}
\end{center}
\caption{Contribution from term 4}
\label{fig:col3}
\vspace*{1cm}
\end{figure}

For last term $C(0,0,0)$, which factorizes in $l$ and $k$ we can
close so that in the $l_1$ integration only $P_2$ contributes,
and in the $k_1$ integration only $P_5$. This results in a square
(\fig{col3}).

Next we will do the $t$ and later the $s$ integration analytically.
Since we left $m_6$ finite, after partial fractioning the integrals
are of the form
\begin{eqnarray}
\int\limits_0^\infty ds \int\limits_0^\infty dt
\frac{1}{s+s_0-i0}\frac{1}{t+t_0-i0}
\frac{1}{\sqrt[c]{(at+b+i0+cs)^2-4st}}
& & \mbox{for term 1}
\nonumber\\
\int\limits_0^\infty ds \int\limits_0^\infty dt
\frac{1}{s+s_0-i0}\frac{1}{t+t_0-i0}
\frac{1}{at+b+i0+cs}
& & \mbox{for term 2 \& 3}
\nonumber\\
\int\limits_0^\infty ds \int\limits_0^\infty dt
\frac{1}{s+s_0-i0}\frac{1}{s+s_0'-i0}
\frac{1}{t+t_0-i0}\frac{1}{t+t_0'-i0}
& & \mbox{for term 4.}
\end{eqnarray}

\begin{figure}[ht]
\begin{center}
\epsfig{file=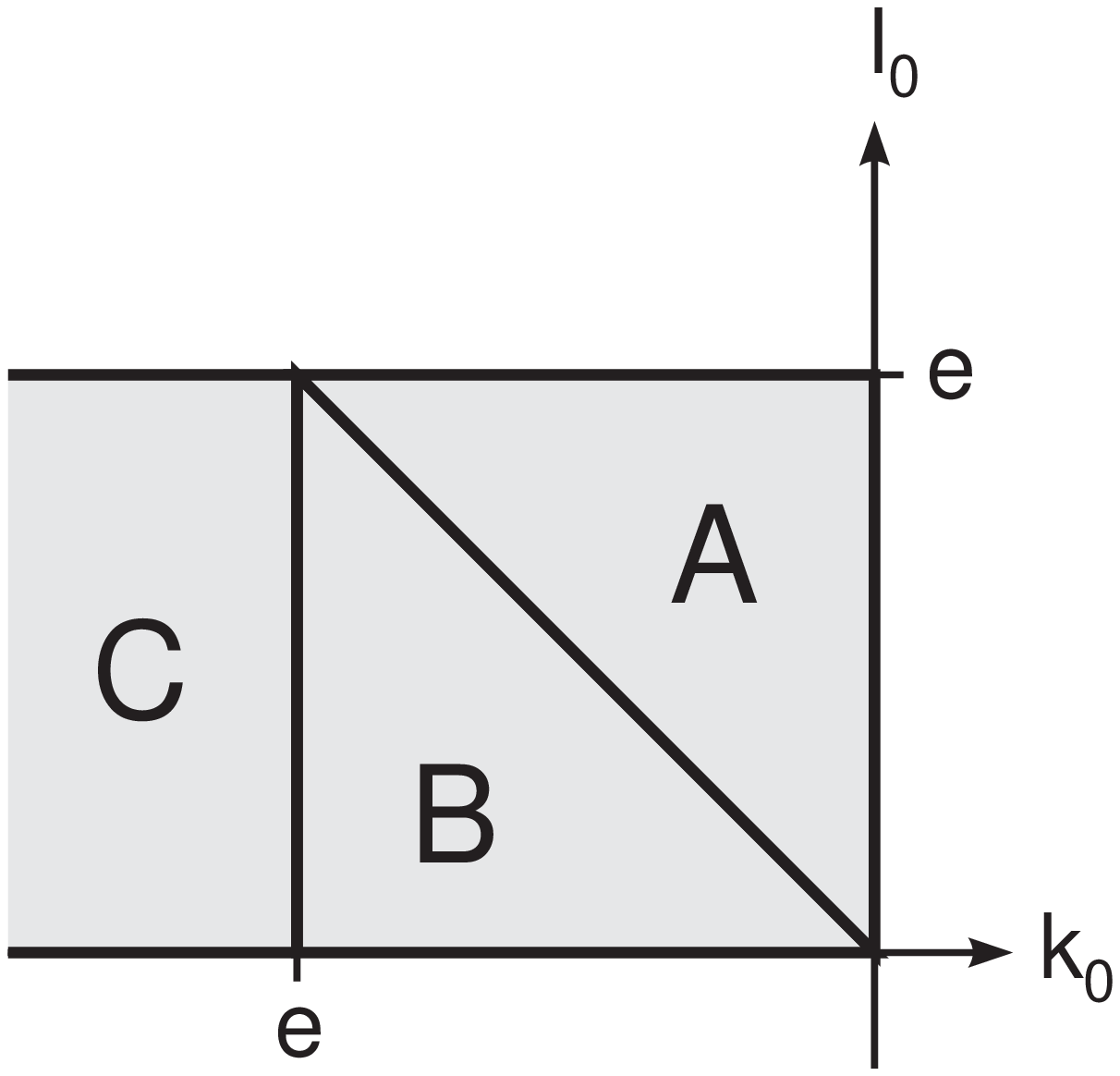,width=0.25\linewidth}
\end{center}
\caption{Integration regions in the $(l_0,k_0)$ plane}
\label{fig:areas}
\vspace*{1cm}
\end{figure}

Several of these integrals become divergent as $m_6 \to 0$,
whenever the corresponding $t_0$ is proportional to $m_6$.
Consequently, the leading term is proportional to $\log(m_6)$.
However, as can be verified, pairs of the above integrals
remain finite in the limit $m_6 \to 0$. The pairs depend
on the position in the $(l_0,k_0)$ plane (area A, B or C,
\fig{areas}).

\section{Numerical Results}
Tables 1--3 summarize the results.
Generally, there are two sets of results
for the momentum expansion method,
namely two $[n/n]$ Pad\'{e} approximants involving $2n+1$ Taylor coefficients,
which document that the results are stable
and converge. It is the second column which should be compared with the
numerical data.

Let us now discuss the three cases separately.
$Case~7$ with equal masses, Table 1,  was the most challenging
for the numerical method. Below the threshold, we could achieve
sufficient numerical accuracy. Above the threshold, the numerical
method could only achieve an accuracy of $\sim 1$\%.
Fortunately, for the equal mass case, M.~Spira could
kindly provide data which were used
for comparison  with the momentum expansion method to high accuracy.
The lack of accuracy in the numerical method in this case is
due to the degenerate cut structure which is present in
the equal mass case. As a result, we are confronted with
two large contributions which almost cancel.

In $Case~7$ with different masses, Table 2, we chose the masses
$m_1=m_2=80 GeV (\sim M_W)$ and $m_6=180GeV (\sim M_t)$ and $\mu =m_1$.
If we wish to calculate the process $Z\rightarrow b\bar{b}$ the
kinematics of interest is $q^2/m_1^2 \sim 1.3$, i.e. far below the
threshold at $q^2/m_1^2=4$. In such a situation the momentum expansion
method always seems to be superior to any other approach (to achieve
10 decimals precision with a few coefficients only is no problem).
Nevertheless, also for low $q^2$ the numerical method yields high
precision as well which will be sufficient for all practical purposes.
Indeed the agreement is quite impressive.

For higher $q^2$ the momentum expansion method naturally
gets less precise, in particular near the thresholds: here
$q^2=4 m_1^2$ and $q^2=(m_1+m_6)^2$, i.e. $q^2/m_1^2 \sim 10.6$.
Between the thresholds and above, both methods again show surprisingly
good agreement.

The situation is indeed quite different for $Case~8$. While for the case
of interest in $Z\rightarrow b\bar{b}$, i.e. $q^2/m_1^2 = 1$ both
approaches yield approximately the same precision, the momentum
expansion method looses quickly for the higher $q^2$, i.e. many
more Taylor coefficients would be needed.

Quite generally speaking the results of the momentum expansion method
obtained so far seem to indicate that the more heavy masses in a diagram
occur, the better the convergence of the Pad\'{e} approximants. This
is at least qualitatively plausible since we start from a large mass
expansion. It is also numerically verified comparing $Case~7$ and 8.

In any case, splitting off systematically zero-mass thresholds yields
more precise results for the momentum expansion method than keeping
small but non-vanishing masses resulting in low thresholds. This has been
discussed in \cite{Rhei} for $Case~5$ (i.e. $m_5=m_6=M$ and all others 0,
see Tables 3 and 4 of that Ref.).
For the numerical method to handle such cases is much easier. In fact,
one could rely on the original integral representations \cite{CKK,FKK}
without the modifications considered here and achieve
much better numerical results for all cases.

Concerning the CPU-time, comparison of the methods is difficult.
In the momentum expansion method the real job is to calculate the Taylor
coefficients. Once they are known, for any $q^2$ (in the complex plane)
the calculation of the diagram under consideration is a matter of seconds.
Thus, providing Taylor coefficients for the diagrams is something which
practically settles the calculation of the diagrams under consideration
once for all in a wide range of $q^2$. This is in particular true if only
one nonzero mass is involved since in this case the Taylor coefficients
are just rational numbers plus some well known irrationals like $\zeta(2)$,
$\zeta(3)$ etc. The methods to calculate Taylor coefficients are at present
still a matter of intense investigation (see e.g. also \cite{TC}) and the
hope is that we will soon have more efficient methods for their calculation
and higher coefficients than used here can be obtained in the future.\\

For the numerical method, we can give a rough estimate by comparison with
the results of \cite{CKK}. Due to the extra subtraction terms, the
CPU time needed here is about an order of magnitude larger than in the 
IR- and collinear convergent cases studied in \cite{CKK}.
\section{Conclusion}

Two different methods to calculate scalar two-loop vertex diagrams with
infrared and collinear divergences have been investigated and
interesting techniques for their calculation have been developed.
An important case with different masses has been solved, which is
a step forward to a realistic calculation of the two-loop decay
$Z\rightarrow b\bar{b}$.
The results show that both methods deliver consistent results.
This confirms the expectation that the two techniques developed
are applicable in the demanding cases considered here.

\bigskip
\noindent
{\large{\bf Acknowledgement}}

\medskip
We are grateful to M.~Spira for providing us with numerical data for
the $Case~7$ with equal masses. J.F. is grateful to O.~Veretin 
and T.~Kotikov for helpful discussions. K.S.
would like to thank Ben Grinstein and the UCSD physics department for their
hospitality.


\newpage

{\scriptsize
\begin{table}
\tcaption{The finite part for $Case~7$ with equal masses.}
\medskip
\def\.{&.}\def\pl{&$\pm$}
\halign{\strut\vrule~\hfil#&#~\vrule
&~\hfil#&#
&~\hfil#&#~\vrule &~\hfil#&# &~\hfil#&#~\vrule  &~\hfil#&# &~\hfil#&#~\vrule
\cr
\noalign{\hrule}
& $q^2/M^2$ && [10/10]  &&&& [14/14]  &&&& numerical results
&& \cr
&&& Re && Im && Re && Im && Re && Im \cr
\noalign{\hrule}
1&.0
& $ $2&.764761286   & $ -$0&.1635351018
& $ $2&.764761286   & $ -$0&.1635351018
& $ $2&.76477       & $ -$0&.16353
\cr
2&.0
& $ $1&.619771096   & $ $0&.4534319337
& $ $1&.619771096   & $ $0&.4534319337
& $ $1&.61977       & $ $0&.45344
\cr
3&.0
& $ $1&.477131552    & $  $0&.5068403285
& $ $1&.477131552    & $  $0&.5068403285
& $ $1&.47713        & $  $0&.50684
\cr
3&.9
& $ $3&.0354569     & $ $0&.28920132
& $ $3&.035456900   & $ $0&.2892013177
& $ $3&.03546       & $ $0&.28920
\cr
3&.99
& $ $5&.1725        & $ $0&.0143
& $ $5&.17259       & $ $0&.014233
& $ $5&.17253       & $ $0&.014236
\cr
4&.01
& $ $6&.188         & $ $2&.748
& $ $6&.187         & $ $2&.749
& $ $6&.18778       & $ $2&.74948
\cr
4&.1
& $ $3&.27648       & $ $3&.92262
& $ $3&.27645       & $ $3&.92263
& $ $3&.27646       & $ $3&.92263
\cr
5&.0
& $ -$0&.438728     & $ $2&.54547
& $ -$0&.438728258  & $ $2&.545472409
& $ -$0&.438728     & $ $2&.54548
\cr
6&.0
& $ -$0&.89277747    & $ $1&.49214654
& $ -$0&.8927774793  & $ $1&.492146568
& $ -$0&.892777      & $ $1&.49215
\cr
7&.0
& $ -$0&.90571711    & $ $0&.92720655
& $ -$0&.9057171008  & $ $0&.9272065484
& $ -$0&.905717      & $ $0&.927207
\cr
8&.0
& $ -$0&.828158487   & $ $0&.59820497
& $ -$0&.8281584836  & $ $0&.5982049870
& $ -$0&.828159      & $ $0&.598205
\cr
10&.0
& $ -$0&.65092293     & $ $0&.26109287
& $ -$0&.6509229304   & $ $0&.2610928562
& $ -$0&.650923       & $ $0&.261093
\cr
40&.0
& $ -$0&.07133439     & $ -$0&.0467805
& $ -$0&.0713343755   & $ -$0&.0467805683
& $ -$0&.0713345      & $ -$0&.0467805
\cr
100&.0
& $ -$0&.0126312     & $ -$0&.0182863
& $ -$0&.012631500   & $ -$0&.018286325
&      &  N/A        &      & N/A
\cr
400&.0
& $ -$0&.000556       & $ -$0&.002743
& $ -$0&.00055586     & $ -$0&.00274078
& $ -$0&.000555895    & $ -$0&.00274078
\cr
\noalign{\hrule}}
\end{table}
}

\vspace{-2.5cm}

{\scriptsize
\begin{table}
\tcaption{The finite part for $Case~7$ with different masses.}
\medskip
\def\.{&.}\def\pl{&$\pm$}
\halign{\strut\vrule~\hfil#&#~\vrule
&~\hfil#&#
&~\hfil#&#~\vrule &~\hfil#&# &~\hfil#&#~\vrule  &~\hfil#&# &~\hfil#&#~\vrule
\cr
\noalign{\hrule}
& $q^2/m_1^2$ && [12/12]  &&&& [15/15]  &&&& numerical results
&& \cr
&&& Re && Im && Re && Im && Re && Im \cr
\noalign{\hrule}
1&.0
& $ $1&.255976034   & $ $0&.4303855857
& $ $1&.255976034   & $ $0&.4303855857
& $ $1&.255974      & $ $0&.430385
\cr
2&.0
& $ $0&.6127661221   & $ $0&.4991658334
& $ $0&.6127661221   & $ $0&.4991658334
& $ $0&.61277        & $ $0&.499164
\cr
3&.0
& $ $0&.4346988621    & $ $0&.4502907526
& $ $0&.4346988621    & $ $0&.4502907526
& $ $0&.43470         & $ $0&.450292
\cr
3&.9
& $ $0&.4583945    & $ $0&.34378450
& $ $0&.458394486  & $ $0&.343784488
& $ $0&.4584       & $ $0&.34379
\cr
3&.99
& $ $0&.4926   & $ $0&.2501
& $ $0&.49254  & $ $0&.25004761
& $ $0&.4925   & $ $0&.25003
\cr
4&.01
& $ $0&.637   & $ $0&.190
& $ $0&.6359  & $ $0&.1902
& $ $0&.63598 & $ $0&.18994
\cr
4&.1
& $ $0&.73506  & $ $0&.43258
& $ $0&.73504  & $ $0&.43260
& $ $0&.73526  & $ $0&.43271
\cr
5&.0
& $ $0&.32211   & $ $0&.855621
& $ $0&.322110  & $ $0&.855622
& $ $0&.32219   & $ $0&.85569
\cr
6&.0
& $ $0&.00752     & $ $0&.802352
& $ $0&.0075200   & $ $0&.8023528
& $ $0&.00756     & $ $0&.80237
\cr
7&.0
& $ -$0&.16178    & $ $0&.682114
& $ -$0&.161777   & $ $0&.682110
& $ -$0&.16176    & $ $0&.68214
\cr
8&.0
& $ -$0&.25256   & $ $0&.56468
& $ -$0&.252547  & $ $0&.564693
& $ -$0&.25255   & $ $0&.56469
\cr
10&.0
& $ -$0&.3259    & $ $0&.3799
& $ -$0&.3256    & $ $0&.37987
& $ -$0&.32559   & $ $0&.37975
\cr
40&.0
& $ -$0&.06956    & $ -$0&.02984
& $ -$0&.0695667  & $ -$0&.029828
& $ -$0&.06956    & $ -$0&.02983
\cr
100&.0
& $ -$0&.01350    & $ -$0&.01529
& $ -$0&.013493   & $ -$0&.015279
& $ -$0&.01349    & $ -$0&.01528
\cr
400&.0
& $ -$0&.00072     & $ -$0&.00253
& $ -$0&.00071     & $ -$0&.002537
& $ -$0&.00071     & $ -$0&.00254
\cr
\noalign{\hrule}}
\end{table}
}

\vspace{-2.5cm}

{\scriptsize
\begin{table}
\tcaption{The finite part for $Case~8$.}
\medskip
\def\.{&.}\def\pl{&$\pm$}
\halign{\strut\vrule~\hfil#&#~\vrule
&~\hfil#&#
&~\hfil#&#~\vrule &~\hfil#&# &~\hfil#&#~\vrule  &~\hfil#&# &~\hfil#&#~\vrule
\cr
\noalign{\hrule}
& $q^2/m_6^2$ && [12/12]  &&&& [15/15]  &&&& numerical results
&& \cr
&&& Re && Im && Re && Im && Re && Im \cr
\noalign{\hrule}
0&.5
& $ $81&.17501719   & $ $12&.06458720
& $ $81&.17501719   & $ $12&.06458720
& $ $81&.1750       & $ $12&.0644
\cr
1&.0
& $ $17&.766     & $ $19&.97799834
& $ $17&.7659    & $ $19&.97799834
& $ $17&.7658    & $ $19&.9779
\cr
2&.0
& $ -$0&.605   & $ $7&.376
& $ -$0&.6047  & $ $7&.3759
& $ -$0&.604   & $ $7&.376
\cr
3&.0
& $ -$1&.753   & $ $3&.182
& $ -$1&.7543  & $ $3&.1815
& $ -$1&.754   & $ $3&.182
\cr
4&.0
& $ -$1&.596    & $ $1&.606
& $ -$1&.595    & $ $1&.603
& $ -$1&.595    & $ $1&.604
\cr
5&.0
& $ -$1&.33     & $ $0&.881
& $ -$1&.327    & $ $0&.880
& $ -$1&.326    & $ $0&.880
\cr
6&.0
& $ -$1&.10     & $ $0&.50
& $ -$1&.096    & $ $0&.503
& $ -$1&.095    & $ $0&.503
\cr
7&.0
& $ -$0&.92      & $ $0&.28
& $ -$0&.914     & $ $0&.289
& $ -$0&.913     & $ $0&.288
\cr
8&.0
& $ -$0&.77  & $ $ 0&.15
& $ -$0&.773 & $ $ 0&.159
& $ -$0&.771 & $ $ 0&.159
\cr
10&.0
& $ -$ 0&.56     & $ $ 0&.01
& $ -$ 0&.572    & $ $ 0&.023
& $ -$ 0&.570    & $ $ 0&.025
\cr
\noalign{\hrule}}
\label{padecut}
\end{table}
}

\newpage

\begin{thebibliography}{99}

\bibitem{ft}
J. Fleischer and O.V. Tarasov, Z. Phys. C 64 (1994) 413.

\bibitem{fst}
J. Fleischer, V.A.~Smirnov and O.V. Tarasov ,
Z. Phys. C74 (1997) 379.

\bibitem{CKK}
A.~Czarnecki, U.~Kilian, D.~Kreimer,  
Nucl.~Phys. B433, 259 (1995).

\bibitem{FKKR}
A.~Frink, B.A.~Kniehl, D.~Kreimer, K.~Riesselmann, 
Phys. Rev. D54 (1996) 4548.

\bibitem{ae}
S.G.~Gorishny, preprints JINR E2--86--176, E2--86--177 (Dubna 1986);
Nucl. Phys. B319 (1989) 633;
K.G.~Chetyrkin, Teor. Mat. Fiz. 75 (1988) 26; 76 (1988) 207;
K.G.~Chetyrkin, preprint MPI-PAE/PTh 13/91 (Munich, 1991);
S.G.~Gorishny and S.A. Larin, Nucl. Phys. B~283 (1987) 452;
S.A. Larin, T. van Ritbergen and J.A.M.~Vermaseren,
Nucl.~Phys. B~438 (1995) 278.

\bibitem{ae-proof}
V.A.~Smirnov,  Commun. Math. Phys. 134 (1990) 109;
V.A.~Smirnov, {\em Renormalization and asymptotic expansions}
(Birkh\"{a}user, Basel, 1991).

\bibitem{vs}
V.A.~Smirnov, Mod. Phys. Lett. A 10 (1995) 1485.

\bibitem{aems}
V.A.~Smirnov, Phys.~Lett. B394 (1997) 205;
A.~Czarnecki and V.A.~Smirnov, Phys.~Lett. B394 (1997) 211;
V.A.~Smirnov, hep-ph/9703357.

\bibitem{Rhei}
J. Fleischer and O.V. Tarasov, Proceedings of the 1966 Zeuthen Workshop
on Elementary Particle Theory: QCD and QED in higher Orders; Rheinsberg
Germany 21-26 April 1966. Nucl.~Phys.~B (Proc. Suppl.) 51C (1966);
J.~Bl\"umlein, F.~Jegerlehner and T.~Riemann editors.

\bibitem{FKK}
A.~Frink, U.~Kilian, D.~Kreimer, 
Nucl.Phys.B488 (1997) 426.

\bibitem{Laus}
L.~Br\"ucher, J.~Franzkowski, A.~Frink, D.~Kreimer, 
`An Introduction to XLOOPS´,
MZ-TH/96-39A-D, hep-ph/9611378, 4 talks,
all to appear in Nucl.Instr.Meth.Phys.Res.A,
Presented at 5th
International Workshop on New Computing Techniques in Physics Research
(AIHENP96), Lausanne, Switzerland, 2-6 Sep 1996.

\bibitem{mpfun}
D.H. Bailey,
{\em ACM Transactions on Mathematical Software}, 19 (1993) 288.

\bibitem{dimreg}
G.~'t Hooft and M.~Veltman,  Nucl.~Phys. B44 (1972) 189;
C.G.~Bollini and J.J.~Giambiagi,  Nuovo Cim. 12B (1972) 20.

\bibitem{ZiF}
J. Fleischer and O.V. Tarasov, in proceedings of the ZiF conference on
{\em Computer Algebra in Science and Engineering},
Bielefeld, 28-31 August 1994, World Scientific 1995, J.~Fleischer,
J.~Grabmeier, F.W.~Hehl and W.~K\"uchlin editors.

\bibitem{FORM} J.A.M.~Vermaseren, {\em Symbolic manipulation with
FORM} (Amsterdam, Computer Algebra Nederland, 1991).

\bibitem{REDUCE}
A.C.~Hearn,
REDUCE {\em User's manual}, version 3.5, RAND publication CP78 (Rev.10/93).

\bibitem{Ber}
F.A. Berends, A.I. Davydychev, V.A. Smirnov and J.B. Tausk,
Nucl. Phys. B439 (1995) 536.

\bibitem{2lp3pt}
D.~Kreimer, 
Phys.~Lett. B292, 341, (1992).

\bibitem{TC}
O.V. Tarasov, Nucl.Phys. B 480 (1996) 397.

\end{thebibliography}
\end{document}